\documentclass[12pt]{article}
\usepackage{amssymb}
\usepackage{amsfonts}
\usepackage{amsmath}
\usepackage{graphicx}

\begin{document}

\title{Knot polynomial invariants in classical Abelian Chern-Simons field
theory}
\author{Xin LIU \\
School of Mathematics and Statistics, University of Sydney\\
NSW 2006, Australia\\
liuxin@maths.usyd.edu.au}
\date{ }
\maketitle

\begin{abstract}
Kauffman knot polynomial invariants are discovered in classical abelian
Chern-Simons field theory. A topological invariant $t^{I\left( \mathcal{L}
\right) }$ is constructed for a link $\mathcal{L}$, where $I$ is the abelian
Chern-Simons action and $t$ a formal constant. For oriented knotted vortex
lines, $t^{I}$ satisfies the skein relations of the Kauffman R-polynomial;
for un-oriented knotted lines, $t^{I}$ satisfies the skein relations of the
Kauffman bracket polynomial. As an example the bracket polynomials of
trefoil knots are computed, and the Jones polynomial is constructed from the
bracket polynomial.

\bigskip


\end{abstract}

\section{Introduction}

\label{Section_1}

Quantum Chern-Simons (CS) theories are one of the most important
three-dimensional topological quantum field theories. Witten discovered that
\cite{WittenCMP1989knots} quantum CS theories provide a natural field
theoretical origin for link invariants, beyond their algebraic origin from
quantum groups \cite{Jones}. Link invariants are the central concept of knot
theory used to classify knot equivalence classes; since three-manifolds are
related to knots via Dehn surgery, link invariants also yield three-manifold
topological invariants. Building on Witten's breakthrough, considerable knot
and three-manifold invariants have been constructed \cite%
{GuadagniniNPB1990,Labastida}. They can be organised using the Kontsevich
integral, and will lead to perturbative CS theories with every term
containing a finite type LMO invariants \cite{LMO}. Recent developments
include the Rozansky-Witten model \cite{RW9612216,GThompson} and the
Gaiotto-Witten-Kapustin-Saulina model \cite{GW08042907}, which are
constructed to act as the Grassmann-odd versions of the CS actions
respectively with normal and super Lie gauge groups. In comparison with CS
theories, in these models a large number of terms are dropped from the CS
perturbative expansions and hence computation is simplified and elementary
information is extracted.

However, for classical CS theories, there are no such direct relationships
between link invariants and CS theories. Classical CS theories are based on
the CS action, which bears different meaning in various physical problems
--- a most important example is the helicity in fluid mechanics. Moffatt
introduced the concept of helicity and revealed its conservation during
evolution of fluid flow \cite{Moffatt1969}. Arnol'd showed that helicity is
invariant under volume-preserving diffeomorphisms \cite{Arnol'd}. Moffatt
and Ricca discovered that for a magnetic fluid containing knotted magnetic
lines of force its helicity can be given by self-linking and linking numbers
of knots \cite{Moffatt1969,MoffattBookChap}. This provides an algebraic
method to count magnetic fluid helicity, much simpler than computation of CS
$3$-form integrals. Today helicity is important in research of knotted
vortex lines in optical beams, Bose-Einstein condensates,
magnetohydrodynamics of solar plasma and so on. However, as mentioned, in
the study of classical CS theories we still need to find direct
relationships between the CS theories and link polynomial invariants, the
powerful tool of knot theory for classification of knot equivalence classes,
as happened in the case of quantum CS theories. In this regard in this paper
we attempt to find polynomial invariants associated to knotted vortex lines
in the framework of classical CS theories.

The abelian Chern-Simons action is given by%
\begin{equation}
I=\frac{1}{4\pi }\int_{M}A\wedge F=\frac{1}{8\pi }\int_{M}\epsilon
^{ijk}A_{i}F_{jk}d^{3}x.  \label{helicity0}
\end{equation}%
Here $A_{i}$ is a $U\left( 1\right) $ gauge potential and $F_{ij}=\partial
_{i}A_{j}-\partial _{j}A_{i}$ the field tensor. In the hydrodynamical
formulism of quantum mechanics, $A_{i}$ is the velocity field distributed
within a quantum fluid, $F_{ij}$ is the vorticity, and $I$ the fluid
helicity up to dimensional constants. $A_{i}$ is defined as $A_{i}=\frac{1}{%
2i\psi ^{\ast }\psi }\left( \psi ^{\ast }\partial _{i}\psi -\partial
_{i}\psi ^{\ast }\psi \right) $ in terms of the complex scalar wave function
$\psi \left( \vec{x}\right) $ describing the physical system, $\psi =\phi
^{1}+i\phi ^{2}$, with $\phi ^{a}\in \mathbb{R}$, $a=1,2$. Defining a
two-dimensional unit vector $n^{a}$ from $\phi ^{a}$, $n^{a}=\frac{\phi ^{a}%
}{\left\Vert \phi \right\Vert }$, the potential $A_{i}$ can be expressed as%
\begin{equation}
A_{i}=\epsilon ^{ab}n^{a}\partial _{i}n^{b}.  \label{U1decomp}
\end{equation}%
The field tensor $F_{ij}$ has a quantum mechanical expression in terms of $%
n^{a}:F_{ij}=2\epsilon ^{ab}\partial _{i}n^{a}\partial _{j}n^{b}$. It can be
proved that \cite{MyPRD} there is a $\delta $-function residing in $F_{ij}:$
$\frac{1}{8\pi }\epsilon ^{ijk}F_{jk}=\delta ^{2}\left( \phi \right)
D^{i}\left( \frac{\phi }{x}\right) $, where $D^{i}\left( \frac{\phi }{x}%
\right) =\frac{1}{2}\epsilon ^{ijk}\epsilon ^{ab}\partial _{j}\phi
^{a}\partial _{k}\phi ^{b}$ is a Jacobian determinant, and $\delta
^{2}\left( \phi \right) $ is the $\delta $-function which does not vanish
only at zero-points of $\phi ^{a}$ (i.e., at singular points of $n^{a}$).
The field tensor $F_{ij}$ is non-trivial only at where the zero-point
equations, $\phi ^{1,2}\left( \vec{x}\right) =0,$ are satisfied. In the
three-dimensional real space the solutions to the two zero-point equations
are a family of, say, $N$ isolated singular line structures. These lines are
just the vortex lines arising from singularity of the field tensor $F_{ij}$.
Let $\xi _{k}$ denote the $k$-th line with a parametric equation $%
x^{i}=z_{k}^{i}\left( s\right) ,\ $where $s$ is the line parameter. Locally,
the unit vector $n^{a}$ lies in the two-dimensional plane normal to $\xi
_{k} $, with the intersection point between $\xi _{k}$ and the plane being
the singular point of the $n^{a}$ field. Then $\delta ^{2}\left( \phi
\right) $ can be expanded onto these $N$ lines as $\delta ^{2}\left( \phi
\right) D^{i}\left( \frac{\phi }{x}\right) =\sum_{k=1}^{N}W_{k}\int_{\xi
_{k}}\frac{dx^{i}}{ds}\delta ^{3}\left( \vec{x}-\vec{z}_{k}\left( s\right)
\right) ds,$ where $W_{k}$ is the topological charge of $\xi _{k}$. In
hydrodynamics $W_{k}$ may carry the meaning of fluid flux. Thus the CS
action becomes $I=\frac{1}{2\pi }\sum_{k=1}^{N}W_{k}\int_{\xi
_{k}}A_{i}dx^{i}.$ Especially, when the vortex lines are $N$ closed knots
forming a link $\mathcal{L}$, the CS action $I$ becomes a sum of integrals
over $\mathcal{L}$:%
\begin{equation}
I\left( \mathcal{L}\right) =\frac{1}{2\pi }\sum_{k=1}^{N}W_{k}\oint_{\xi
_{k}}A_{i}dx^{i}.  \label{CS1}
\end{equation}

In \cite{MyPRD} we analysed the CS action by means of gauge potential
decomposition and showed that $I$ is closely related to (self-)linkage of
the knots of $\mathcal{L}$:%
\begin{equation}
I\left( \mathcal{L}\right) =\sum_{k=1}^{N}W_{k}^{2}Sl\left( \xi _{k}\right)
+\sum_{k,l=1;k\neq l}^{N}W_{k}W_{l}Lk\left( \xi _{k},\xi _{l}\right) ,
\label{SelfLinklink}
\end{equation}%
where $Sl\left( \xi _{k}\right) $ is the self-linking number of one knot $%
\xi _{k}$, and $Lk\left( \xi _{k},\xi _{l}\right) $ the linking number
between two knots $\xi _{k}$ and $\xi _{l}$.\ Eq.(\ref{SelfLinklink}) is
consistent with the conclusion of Moffatt et al. \cite{MoffattBookChap}.

For the discussions in the following sections it is worth here
having a quick revisit to our analysis of \cite{MyPRD} for
(\ref{CS1}) and (\ref {SelfLinklink}), as follows. Introduce a
$3$-dimensional unit vector $\vec{m}
=\frac{\vec{\kappa}-\vec{\zeta}}{\left\Vert \vec{\kappa}-\vec{\zeta}%
\right\Vert },$ where $\vec{\kappa}$ and $\vec{\zeta}$ are two points
respectively picked up from two knots $\xi _{k}$ and $\xi _{l}$ of the link $%
\mathcal{L}$. When $\vec{\kappa}$ and $\vec{\zeta}$ run along knots of $%
\mathcal{L}$, the $\vec{m}$ runs over the $2$-dimensional sphere $S^{2}$ in
the $3$-dimensional space. On this $S^{2}$ we introduce a unit vector $e^{A}$%
, with $A=1,2$ denoting the local coordinates on $S^{2}$. Apparently $\vec{e}
$ is always perpendicular to $\vec{m}$. In \cite{MyPRD} the $\vec{e}$ is
used to re-express the gauge potential $A_{i}$ as $A_{i}=\epsilon
^{AB}e^{A}\partial _{i}e^{B}$, hence (\ref{CS1}) becomes $I\left( \mathcal{L}%
\right) =\frac{1}{2\pi }\sum_{k=1}^{N}W_{k}\oint_{\xi _{k}}\epsilon
^{AB}e^{A}\partial _{i}e^{B}d\kappa ^{i}$. With the consideration that $%
m^{i} $ is defined from both $\xi _{k}$ and $\xi _{l}$, we write $I\left(
\mathcal{L}\right) $ more symmetrically as%
\begin{equation}
I\left( \mathcal{L}\right) =\frac{1}{4\pi }\sum_{k,l=1}^{N}W_{k}W_{l}\oint_{%
\xi _{k}}d\kappa ^{i}\oint_{\xi _{l}}d\zeta ^{j}\epsilon ^{AB}\partial
_{i}e^{A}\partial _{j}e^{B}.  \label{CS2}
\end{equation}%
Now let us investigate (\ref{CS2}) by examining the two points $\vec{\kappa}$
and $\vec{\zeta}$. Eq.(\ref{CS2}) contains three cases in regard to
different relative positions of $\vec{\kappa}$ and $\vec{\zeta}$: (i) $\xi
_{k}$ and $\xi _{l}$ are different knots and $\vec{\kappa}$ and $\vec{\zeta}$
are different points; (ii) $\xi _{k}$ and $\xi _{l}$ are a same knot but $%
\vec{\kappa}$ and $\vec{\zeta}$ are different points; (iii) $\vec{\kappa}$
and $\vec{\zeta}$ are a same point.

\begin{itemize}
\item Case (i): In regard to the definitions of $\vec{m}$ and $\vec{e}$, we
see that $\epsilon ^{AB}e^{A}\partial _{i}e^{B}$ indeed gives a Wu-Yang
potential \cite{Wu-Yang}. Hence (\ref{CS2}) leads to $I\left( \mathcal{L}%
\right) _{\mathrm{(i)}}=\sum_{k,l=1;k\neq l}^{N}\frac{W_{k}W_{l}}{4\pi }%
\oint_{\xi _{k}}d\kappa ^{i}\oint_{\xi _{l}}d\zeta ^{j}\left( \vec{m}\cdot
\partial _{i}\vec{m}\times \partial _{j}\vec{m}\right) $, where $\vec{m}%
\cdot \partial _{i}\vec{m}\times \partial _{j}\vec{m}$ is recognized to be a
surface element of the $S^{2}$. According to \cite{Polyakov}, $I\left(
\mathcal{L}\right) _{\mathrm{(i)}}$ presents the linking number between $\xi
_{k}$ and $\xi _{l}:I\left( \mathcal{L}\right) _{\mathrm{(i)}%
}=\sum_{k,l=1;k\neq l}^{N}W_{k}W_{l}Lk\left( \xi _{k},\xi _{l}\right) $.

\item Case (ii): Similarly, the integral of (\ref{CS2}) leads to the
writhing number $Wr\left( \xi _{k}\right) $ of the knot $\xi _{k}$: $I\left(
\mathcal{L}\right) _{\mathrm{(ii)}}=\sum_{k=1}^{N}W_{k}^{2}Wr\left( \xi
_{k}\right) $.

\item Case (iii): In this case $\vec{m}$ becomes the tangent vector $\vec{T}$
of the vortex line $\xi _{k}$. And $\vec{e}$ becomes the vector normal to $%
\vec{T}$, having arbitrariness of rotating about $\vec{T}$. On the one hand,
according to differential geometry of curves, $\xi _{k}$ possesses a Frenet
frame formed by three orthonormal vectors: $\vec{T}$, $\vec{N}$ and $\vec{B}$%
, where $\vec{N}$ and $\vec{B}$ are the so-called normal and bi-normal unit
vectors. On the other hand, for the decomposition (\ref{U1decomp}) of the $%
U\left( 1\right) $ potential $A_{i}$, on a plane normally intersecting $\xi
_{k}$ the intersection point is a singular ill-defined point of the $\vec{n}$
field. Then, noticing $\vec{n}$ is in the same plane containing $\vec{e}$,
the singularity of the $\vec{n}$ field can be removed by redefining on $\xi
_{k}$
\begin{equation}
\vec{n}\equiv \vec{e}\equiv \vec{N}.  \label{naredefinitn}
\end{equation}%
Then the double integral (\ref{CS2}) reduces to a single integral $I\left(
\mathcal{L}\right) _{\mathrm{(iii)}}$ $=\frac{1}{2\pi }%
\sum_{k=1}^{N}W_{k}^{2}\oint_{\xi _{k}}\epsilon ^{ab}n^{a}\partial
_{i}n^{b}d\kappa ^{i}$ $=\frac{1}{2\pi }\sum_{k=1}^{N}W_{k}^{2}Tw\left( \xi
_{k}\right) $, yielding the twisting number of $\xi _{k}$, $Tw\left( \xi
_{k}\right) =\oint_{\xi _{k}}\vec{N}\cdot d\vec{B}$.
\end{itemize}

\noindent Thus, in the light of the Calugareanu-White formula $Sl\left( \xi
_{k}\right) =Wr\left( \xi _{k}\right) +Tw\left( \xi _{k}\right) $, one can
summarize Cases (i)--(iii) to obtain the above result (\ref{SelfLinklink}):%
\begin{equation}
I\left( \mathcal{L}\right) =I\left( \mathcal{L}\right) _{\mathrm{(i)}%
}+I\left( \mathcal{L}\right) _{\mathrm{(ii)}}+I\left( \mathcal{L}\right) _{%
\mathrm{(iii)}}=\sum_{k=1}^{N}W_{k}^{2}Sl\left( \xi _{k}\right)
+\sum_{k,l=1;k\neq l}^{N}W_{k}W_{l}Lk\left( \xi _{k},\xi _{l}\right) .
\label{CS3}
\end{equation}%
Here a point should be emphasized. It is seen that in Cases (i) and (ii) the
points $\vec{\kappa}$ and $\vec{\zeta}$ are different, no matter the knots $%
\xi _{k}$ and $\xi _{l}$ are different or not. Therefore, Case (i) plus (ii)
indeed wrap up all the contributions of \textquotedblleft
different-point-defined $\vec{m}$\textquotedblright\ to $I\left( \mathcal{L}%
\right) $. This fact will play an important role in the next section.

In this paper we will further study deeper algebraic essence of $I\left(
\mathcal{L}\right)$ and reveal its relationship to the polynomial invariants
of knot theory. Significance of this study dwells in that it establishes a
bridge between the CS action and algebraic polynomial invariants of knot
theory. The paper is arranged as follows. In Section \ref{Section_2} the
skein relations of the Kauffman R-polynomial for oriented knotted vortex
lines will be obtained. In Section \ref{Section_3} the skein relations of
the Kauffman bracket polynomial for un-oriented knotted lines will be
obtained. As an example the bracket polynomials of trefoil knots will be
computed, and the well-known Jones polynomial will be constructed from the
bracket polynomial. Our emphasis is to be placed on Section \ref{Section_3}.
In Section \ref{Section_4} the paper will be summarized and discussions be
presented.

Before proceeding, a preparation should be addressed. Since crossing and
writhing of vortex lines are to be discussed below, vortex lines should have
same topological charges, otherwise the discussion cannot be conducted.
Hence in this paper all topological charges of vortex lines take a same
value: $W_{1}=\cdots =W_{N}=W$. For convenience, one evaluates $W=1$.

\section{Kauffman R-polynomial invariant for oriented knots}

\label{Section_2}

We argue that the exponential%
\begin{equation}
t^{H\left( \mathcal{L}\right) }=t^{\frac{1}{2\pi }\sum_{k}\oint_{\xi
_{k}}A_{i}dx^{i}}  \label{knotinv0}
\end{equation}%
is capable to present the Kauffman R-polynomial for oriented knotted vortex
lines and present the Kauffman bracket polynomial for un-oriented knotted
vortex lines. Here $t$ is a constant, which will appear formally in the
following deduction and may be determined when compared to, say, a concrete
fluid mechanical model. If the theory of this paper could be applied in
another physical problem, which possesses the Chern-Simons-type action (\ref%
{helicity0}), then $t$ would bear a different physical meaning in that
circumstance.

In this section the R-polynomial for oriented knots will be studied.
Oriented knots are useful in solving some physical problems. For instance,
the tangled open vortex lines in Figure 1 can be conveniently studied if
they are regarded as oriented knots:

\begin{figure}[th]
\centering \includegraphics[scale=0.8]{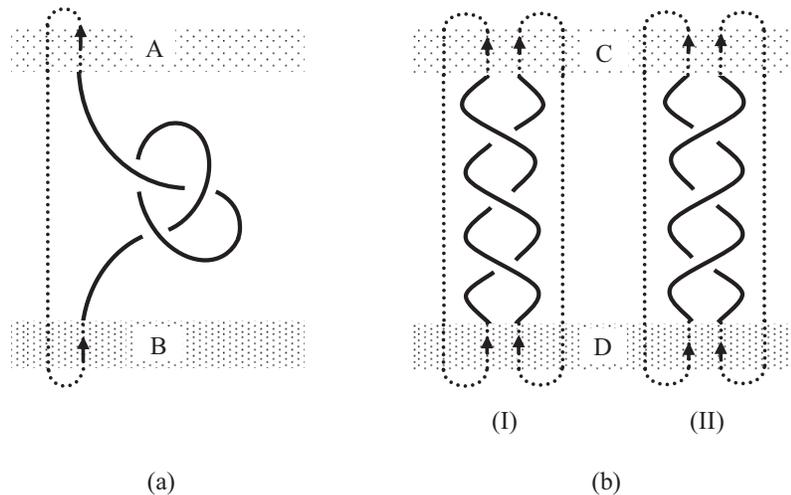}
\caption{Tangled open vortex lines can be studied as oriented knots.}
\label{figure_01}
\end{figure}

\begin{itemize}
\item In Figure \ref{figure_01}(a), to study the tangled line: firstly, one
can extend the open ends at the different boundaries $A$ and $B$ to
infinity, and trivially connect them with the dashed curve to form a closed
loop; secondly, to distinguish $A$ and $B$ a convenient way is to endow the
line with an orientation. Thus the open vortex line can be studied as an
oriented knot;

\item In Figure \ref{figure_01}(b), consider two braids of open vortex lines
which are in different intertwining configurations, respectively marked as
(I) and (II). To distinguish them a reasonable way is: firstly, in each
braid, to connect the open ends at the different boundaries $C$ and $D$ with
the dashed curves shown; secondly, to endow each line with an orientation to
distinguish $C$ and $D$. Thus, the different tangles (I)\ and (II) can be
studied as two oriented knots.
\end{itemize}

\begin{figure}[th]
\centering \includegraphics[scale=0.8]{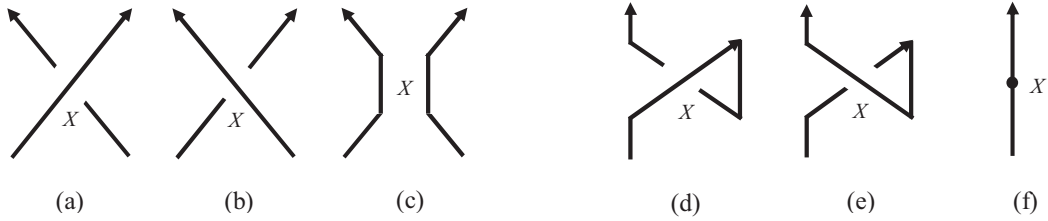}
\caption{Crossing configurations of oriented knots: (a) $l_{+}$; (b) $l_{-}$%
; (c) $l_{0}$. Writhing configurations of oriented knots: (d) $\hat{l}_{+}$;
(e) $\hat{l}_{-}$; (f) $\hat{l}_{0}$.}
\label{figure_02}
\end{figure}

For the purpose of obtaining the R-polynomial from $t^{I\left( \mathcal{L}%
\right) }$, crossing and writhing configurations of links should be studied:
for crossing, consider three links which are almost the same except at one
particular point where different crossing situations occur, as shown in
Figure \ref{figure_02}(a) to \ref{figure_02}(c). The very point $X$ is
called a double point, and the over-crossing, under-crossing and
non-crossing links are respectively denoted by $l_{+},\ l_{-}$ and $l_{0}$;
for writhing, one uses $\hat{l}_{+},\ \hat{l}_{-}$ and $\hat{l}_{0}$ to
denote three links which are almost the same except for different writhing
situations at the point $X$, as shown in Figure \ref{figure_02}(d) to \ref%
{figure_02}(f).

\begin{figure}[th]
\centering \includegraphics[scale=0.8]{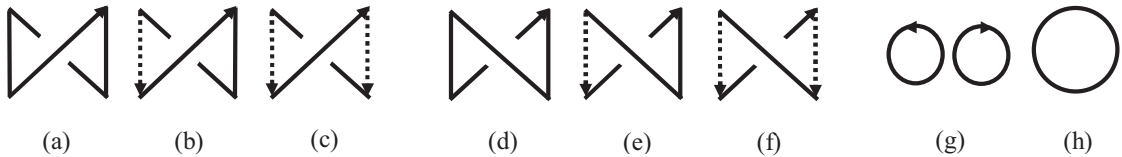}
\caption{Configurations used in the study of oriented knots: (a) $\protect%
\gamma _{+}$; (b) $\protect\gamma _{+}^{\prime }$, containing one
imaginarily added segment; (c) $\protect\gamma _{+}^{\prime \prime
}$, containing two
imaginarily added segments; (d) $\protect\gamma _{-}$; (e) $\protect\gamma %
_{-}^{\prime }$, containing one imaginarily added segment; (f) $\protect\gamma %
_{-}^{\prime \prime }$, containing two imaginarily added segments;
(g) $l_{cc}$; (h) $L_{c}$.} \label{figure_03}
\end{figure}

Now let us examine $t^{I\left( \hat{l}_{+}\right) }$, $t^{I\left( \hat{l}%
_{-}\right) }$ and $t^{I\left( \hat{l}_{0}\right) }$. Since the vortex lines
are oriented, the integration paths can be re-expressed as:
\begin{equation}
\hat{l}_{+}=\hat{l}_{0}\oplus \gamma _{+}^{\phantom{.}\prime }=\hat{l}%
_{0}\oplus \gamma _{+},\ \ \ \ \ \ \ \ \ \hat{l}_{-}=\hat{l}_{0}\oplus
\gamma _{-}^{\phantom{.}\prime }=\hat{l}_{0}\oplus \gamma _{-},
\label{split0}
\end{equation}%
where the symbol \textquotedblleft $\oplus $\textquotedblright\ means
\textquotedblleft union after imaginarily adding and subtracting
paths\textquotedblright , and $\gamma _{+},\ \gamma _{+}^{\phantom{.}\prime
},\ \gamma _{-}$ and $\gamma _{-}^{\phantom{.}\prime }$ are shown in Figure %
\ref{figure_03}(a), \ref{figure_03}(b), \ref{figure_03}(d) and \ref%
{figure_03}(e) respectively. For these imaginary paths we require that
\begin{equation}
t^{I\left( \hat{l}_{\pm }\right) }=t^{I\left( \hat{l}_{0}\oplus \gamma _{\pm
}\right) }=t^{I\left( \gamma _{\pm }\right) }t^{I\left( \hat{l}_{0}\right) },
\label{KaufR0}
\end{equation}%
which demonstrates the difference between the writhing $\hat{l}_{\pm }$ and
the non-writhing $\hat{l}_{0}$.

Reasonability of Eq.(\ref{KaufR0}) is as follows:

\begin{itemize}
\item For (\ref{split0}) let us examine $I\left( \hat{l}_{+}\right) _{%
\mathrm{(i)}}$ and $I\left( \hat{l}_{+}\right) _{\mathrm{(ii)}}$ with
respect to (\ref{CS3}). Noticing that in (\ref{CS3}) the $I\left( \mathcal{L}%
\right) _{\mathrm{(i)}}$ and $I\left( \mathcal{L}\right) _{\mathrm{(ii)}}$
are the contributions of \textquotedblleft different-point-defined $\vec{m}$%
\textquotedblright\ to $I\left( \mathcal{L}\right) $, we pick up two
arbitrary points $\vec{\kappa}$ and $\vec{\zeta}$ from the knots of $\hat{l}%
_{+}$. When doing so, we have three choices:\newline
(1) $\vec{\kappa}$ and $\vec{\zeta}$ both from $\hat{l}_{0}$,\newline
(2) (with loss of generality) $\vec{\kappa}$ from $\hat{l}_{0}$ but $\vec{%
\zeta}$ from $\gamma _{+}$, and\newline
(3) $\vec{\kappa}$ and $\vec{\zeta}$ both from $\gamma _{+}$.\newline
Choice (1) gives $I\left( \hat{l}_{0}\right) _{\mathrm{(i)}}$ and $I\left(
\hat{l}_{0}\right) _{\mathrm{(ii)}}$, which are completely independent of $%
\gamma _{+}$. Choice (2) contributes zero, because $\gamma _{+}$ is isolated
from $\hat{l}_{0}$ without linkage. For Choice (3), only $I\left( \gamma
_{+}\right) _{\mathrm{(ii)}}$ exists because $\gamma _{+}$ is a single knot,
and so Choice (3) yields $Wr\left( \gamma _{+}\right) $. Therefore, Choices
(1)--(3) show complete separation: $I\left( \hat{l}_{+}\right) _{\mathrm{(i)}%
}+I\left( \hat{l}_{+}\right) _{\mathrm{(ii)}}=\left[ I\left( \hat{l}%
_{0}\right) _{\mathrm{(i)}}+I\left( \hat{l}_{0}\right) _{\mathrm{(ii)}}%
\right] +Wr\left( \gamma _{+}\right) $.

\item The $I\left( \mathcal{L}\right) _{\mathrm{(iii)}}$ in (\ref{CS3}) is
\textquotedblleft locally\textquotedblright\ defined, because the vectors $%
\vec{T},\ \vec{N}$ and $\vec{B}$ are locally defined, not coming from
\textquotedblleft different-point-defined $\vec{m}$\textquotedblright .
Hence when the path $\hat{l}_{+}$ turns into $\hat{l}_{0}\oplus \gamma _{+}$
in (\ref{split0}), $I\left( \hat{l}_{+}\right) _{\mathrm{(iii)}}$ is
naturally separated as $I\left( \hat{l}_{+}\right) _{\mathrm{(iii)}}=I\left(
\hat{l}_{0}\right) _{\mathrm{(iii)}}+Tw\left( \gamma _{+}\right) $.

\item Therefore, $I\left( \hat{l}_{+}\right) =I\left( \hat{l}_{+}\right) _{%
\mathrm{(i)}}+I\left( \hat{l}_{+}\right) _{\mathrm{(ii)}}+I\left( \hat{l}%
_{+}\right) _{\mathrm{(iii)}}=I\left( \hat{l}_{0}\right) +Sl\left( \gamma
_{+}\right) $, with $Sl\left( \gamma _{+}\right) =I\left( \gamma _{+}\right)
$. Similarly, $I\left( \hat{l}_{-}\right) =I\left( \hat{l}_{0}\right)
+Sl\left( \gamma _{-}\right) $, with $Sl\left( \gamma _{-}\right) =I\left(
\gamma _{-}\right) $.
\end{itemize}

The evaluation of $t^{Sl\left( \gamma _{\pm }\right) }$ is obtained by
computing $Wr\left( \gamma _{\pm }\right) $ and $Tw\left( \gamma _{\pm
}\right) $:%
\begin{equation}
t^{I\left( \gamma _{\pm }\right) }=t^{Sl\left( \gamma _{\pm }\right)
}=t^{\pm \frac{1}{2}}.  \label{selflinknum0}
\end{equation}%
Eq.(\ref{selflinknum0}) is consistent with the algebraically topological
definitions of the self-linking numbers of $\gamma _{\pm }$ \cite%
{Rolfsen,Kauffmanbook}: $Sl\left( \gamma _{\pm }\right) =Sl\left(
L_{c}\right) +\frac{1}{2}\epsilon \left( \gamma _{\pm }\right) =\pm \frac{1}{%
2},$ where $L_{c}$ is a trivial circle shown in Figure \ref{figure_03}(h),
with $Sl\left( L_{c}\right) =0$. And $\epsilon \left( \gamma _{+,-}\right)
=1,-1$ are respectively the degrees of the crossing points of $\gamma _{+,-}$
in Figure \ref{figure_03}(a) and \ref{figure_03}(d).

Thus (\ref{KaufR0}) becomes%
\begin{equation}
t^{I\left( \hat{l}_{\pm }\right) }=t^{\pm \frac{1}{2}}t^{I\left( \hat{l}%
_{0}\right) }.  \label{MyKaufR1}
\end{equation}%
Defining a constant $\hat{\alpha}=t^{I\left( \gamma _{+}\right) }$ [namely $%
\hat{\alpha}^{-1}=t^{I\left( \gamma _{-}\right) }$], and using $R\left(
\mathcal{L}\right) $ to denote $t^{I\left( \mathcal{L}\right) }$, Eq.(\ref%
{MyKaufR1}) gives $R\left( \hat{l}_{+}\right) =\hat{\alpha}R\left( \hat{l}%
_{0}\right) $ and $R\left( \hat{l}_{-}\right) =\hat{\alpha}^{-1}R\left( \hat{%
l}_{0}\right) $, which are known as the second skein relation of the
R-polynomial \cite{Kauffmanbook2}. Furthermore, noticing that $Tw\left(
L_{c}\right) =0$, one obtains $R\left( L_{c}\right) =t^{I\left( L_{c}\right)
}=1$, which is known as the first skein relation of the R-polynomial.

The third skein relation of the R-polynomial reads $R\left( l_{+}\right)
-R\left( l_{-}\right) =zR\left( l_{0}\right) $, where $z$ is a constant. To
obtain this relation the trick \textquotedblleft adding and subtracting
paths\textquotedblright\ could be used again for $l_{\pm }$:%
\begin{equation}
l_{\pm }=l_{0}\oplus \gamma _{\pm }^{\phantom{.}\prime \prime }=l_{0}\oplus
\gamma _{\pm },
\end{equation}%
where $\gamma _{+}^{\phantom{.}\prime \prime }$ and $\gamma _{-}^{\phantom{.}%
\prime \prime }$ are respectively shown in Figure \ref{figure_03}(c) and \ref%
{figure_03}(f). Thus $t^{I\left( l_{+}\right) }=t^{I\left( l_{0}\right)
}t^{I\left( \gamma _{+}\right) }=\hat{\alpha}t^{I\left( l_{0}\right) }$ and $%
t^{I\left( l_{-}\right) }=t^{I\left( l_{0}\right) }t^{I\left( \gamma
_{-}\right) }=\hat{\alpha}^{-1}t^{I\left( l_{0}\right) }$, and%
\begin{equation}
t^{I\left( l_{+}\right) }-t^{I\left( l_{-}\right) }=\left( \hat{\alpha}-\hat{%
\alpha}^{-1}\right) t^{I\left( l_{0}\right) }.  \label{MyKaufR2}
\end{equation}%
Letting $z$ take the value $\hat{\alpha}-\hat{\alpha}^{-1}$, Eq.(\ref%
{MyKaufR2}) gives the relation $R\left( l_{+}\right) -R\left( l_{-}\right)
=zR\left( l_{0}\right) $.

Therefore, in summary, with the definition
\begin{equation}
R\left( \mathcal{L}\right) \equiv t^{I\left( \mathcal{L}\right) },\ \ \ \
\hat{\alpha}\equiv t^{I\left( \gamma _{+}\right) }=t^{\frac{1}{2}},\ \ \ \
z\equiv \hat{\alpha}-\hat{\alpha}^{-1}  \label{orientdenote}
\end{equation}%
for a link $\mathcal{L}$ of oriented knots, we have obtained from $%
t^{I\left( \mathcal{L}\right) }$ the Kauffman R-polynomial invariant $%
R\left( \mathcal{L}\right) $ that satisfies the following three skein
relations \cite{Kauffmanbook2}:%
\begin{eqnarray}
&&R\left( L_{c}\right) =1,  \label{myKaufmanRc} \\
&&R\left( \hat{l}_{+}\right) =\hat{\alpha}R\left( \hat{l}_{0}\right) ,\ \ \
\ R\left( \hat{l}_{-}\right) =\hat{\alpha}^{-1}R\left( \hat{l}_{0}\right) ,
\label{myKaufmanR1} \\
&&R\left( l_{+}\right) -R\left( l_{-}\right) =zR\left( l_{0}\right) .
\label{myKaufmanR2}
\end{eqnarray}%
Kauffman proposed a constant to characterize the R-polynomial: $\delta =%
\frac{\hat{\alpha}-\hat{\alpha}^{-1}}{z}$. Our realization of the Kauffman
R-polynomial corresponds to $\delta =1$.

As an example let us check $R\left( l_{cc}\right) $, where $l_{cc}$ is the
union of two trivial circles, as shown in Figure \ref{figure_03}(g). From (%
\ref{myKaufmanR1}) there are $R\left( \gamma _{+}\right) =\hat{\alpha}%
R\left( L_{c}\right) $ and $R\left( \gamma _{-}\right) =\hat{\alpha}%
^{-1}R\left( L_{c}\right) $, hence in light of (\ref{myKaufmanRc}) one has $%
R\left( \gamma _{+}\right) -R\left( \gamma _{-}\right) =\hat{\alpha}-\hat{%
\alpha}^{-1}$. On the other hand from (\ref{myKaufmanR2}) there is $R\left(
\gamma _{+}\right) -R\left( \gamma _{-}\right) =zR\left( l_{cc}\right) $.
Comparing these two results one obtains $R\left( l_{cc}\right) =\frac{\hat{%
\alpha}-\hat{\alpha}^{-1}}{z}=1.$

\section{Kauffman bracket polynomial invariant for un-oriented knots}

\label{Section_3}

In Eq.(\ref{CS1}) the integration paths have no preferred orientations;
generally, in fluid mechanics and other physical problems the studied closed
vortex lines are un-oriented. Hence, it is natural not to endow closed loops
with orientations when dealing with (\ref{CS1}). In this section we will
show that the CS action induced $t^{I}$ can present the Kauffman bracket
polynomial invariant for un-oriented knots.

\begin{figure}[th]
\centering \includegraphics[scale=0.8]{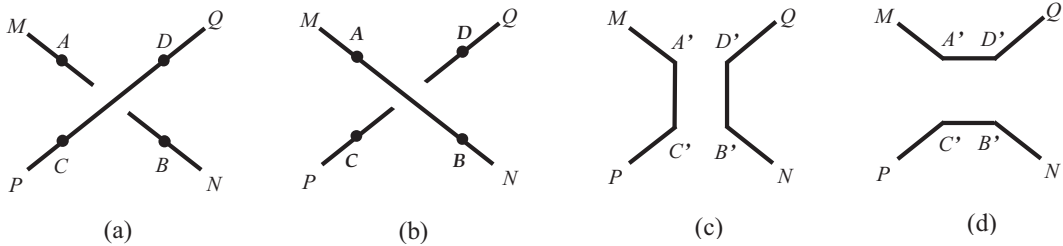}
\caption{Crossing configurations of non-oriented knots: (a) over-crossing $%
L_{+}$; (b) under-crossing $L_{-}$; (c) non-crossing $L_{0}$; (d)
non-crossing $L_{\infty }$.}
\label{figure_04}
\end{figure}

Let $\left\langle \mathcal{L}\right\rangle $ denote the Kauffman bracket
polynomial of a link $\mathcal{L}$ of un-oriented knots. The bracket
polynomial satisfies three skein relations \cite{Kauffmanbook,Kauffmanbook2}%
:
\begin{eqnarray}
&&\left\langle L_{c}\right\rangle =1,  \label{Kaufb1} \\
&&\left\langle L_{+}\right\rangle =a\left\langle L_{0}\right\rangle
+a^{-1}\left\langle L_{\infty }\right\rangle ,\ \ \ \ \ \mathrm{i.e.,}\ \ \
\ \ \left\langle L_{-}\right\rangle =a^{-1}\left\langle L_{0}\right\rangle
+a\left\langle L_{\infty }\right\rangle ,  \label{Kaufb2} \\
&&\left\langle L_{c}\sqcup L\right\rangle =-\left( a^{2}+a^{-2}\right)
\left\langle L\right\rangle .  \label{Kaufb3}
\end{eqnarray}%
Here $a$ is a real constant, $L$ an arbitrary link, and $L_{+}$, $L_{-}$, $%
L_{0}$ and $L_{\infty }$ are crossing and non-crossing configurations shown
in Figure \ref{figure_04}(a) to \ref{figure_04}(d). The symbol
\textquotedblleft $\sqcup $\textquotedblright\ means \textquotedblleft
disjoint union\textquotedblright ; \textquotedblleft $\sqcup $%
\textquotedblright\ is different from the \textquotedblleft $\oplus $%
\textquotedblright\ of the last section, where the former refers to a union
of realistic separate components of a link, while the latter refers to
imaginarily added or subtracted paths.

\noindent Constructing
\begin{equation}
\left\langle \mathcal{L}\right\rangle \equiv t^{I\left( \mathcal{L}\right) },
\end{equation}%
our task is to show that $\left\langle \mathcal{L}\right\rangle $ satisfies (%
\ref{Kaufb1}) to (\ref{Kaufb3}). The first relation (\ref{Kaufb1}) is
satisfied because $Tw\left( L_{c}\right) =0$ and thus $\left\langle
L_{c}\right\rangle =t^{I\left( L_{c}\right) }=1$. For the second and third
relations (\ref{Kaufb2}) and (\ref{Kaufb3}), their verifications will be
detailed respectively in Subsections \ref{subsect3.1} and \ref{subsect3.2},
where the evaluation of the constant $a$ is to be determined. Then, in
Subsection \ref{subsect3.3}, as an example the bracket polynomial for the
right- and left-handed trefoil knots will be computed. In Subsection \ref%
{subsect3.4} the relationship between the Kauffman bracket polynomial and
the Jones polynomial for oriented knots will be given.

\subsection{Skein relation (\protect\ref{Kaufb2})}

\label{subsect3.1}

To realize (\ref{Kaufb2}) the relationships between $L_{\pm }$ and the $%
L_{0} $ and $L_{\infty }$ should be found. For this purpose, as before, we
appeal to the trick \textquotedblleft imaginarily adding
paths\textquotedblright :
\begin{figure}[th]
\centering \includegraphics[scale=0.8]{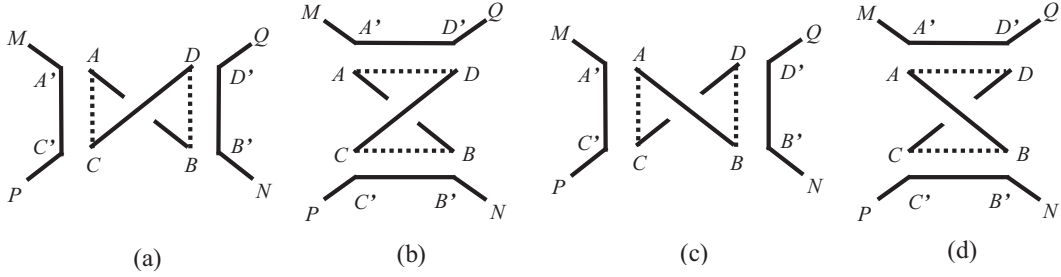}
\caption{Imaginarily adding paths to $L_{+}$ and $L_{-}$: (a) $L_{0}$-type
splitting of $L_{+}$; (b) $L_{\infty }$-type splitting of $L_{+}$; (c) $%
L_{0} $-type splitting of $L_{-}$; (d) $L_{\infty }$-type splitting of $%
L_{-} $.}
\label{figure_05}
\end{figure}

\begin{itemize}
\item To relate $L_{+}$ to $L_{0}$, Figure \ref{figure_04}(a) and \ref%
{figure_05}(a) are considered. On the line $\overline{MN}$ of Figure \ref%
{figure_04}(a), one breaks the point $A$ into a pair $\left( A,A^{\prime
}\right) $, and breaks $B$ into $\left( B,B^{\prime }\right) $, as in Figure %
\ref{figure_05}(a). On the line $\overline{PQ}$ of Figure \ref{figure_04}%
(a), $C$ is broken into $\left( C,C^{\prime }\right) $ and $D$ into $\left(
D,D^{\prime }\right) $ as in Figure \ref{figure_05}(a). Thus $\overline{MN}$
turns to be $\overline{MA^{\prime }}\oplus \overline{AB}\oplus \overline{%
B^{\prime }N}$, and $\overline{PQ}$ to be $\overline{PC^{\prime }}\oplus
\overline{CD}\oplus \overline{D^{\prime }Q}$. Introducing four imaginary
segments $\overline{A^{\prime }C^{\prime }}$, $\overline{AC}$, $\overline{DB}
$ and $\overline{D^{\prime }B^{\prime }}$ to Figure \ref{figure_05}(a), the
two sets $\left( \overline{MA^{\prime }},\overline{A^{\prime }C^{\prime }},%
\overline{C^{\prime }P}\right) $ and $\left( \overline{QD^{\prime }},%
\overline{D^{\prime }B^{\prime }},\overline{B^{\prime }N}\right) $ form an $%
L_{0}$ (i.e. $\overline{MA^{\prime }C^{\prime }P}\oplus \overline{QD^{\prime
}B^{\prime }N}$), while the set $\left( \overline{AC},\overline{CD},%
\overline{DB},\overline{BA}\right) $ forms a writhe $\overline{ACDBA}$ which
is the same as the $\gamma _{+}$ of Figure \ref{figure_03}(a), disregarding
orientations. Let this imaginarily constructed \textquotedblleft $\gamma
_{+} $\textquotedblright\ be denoted by $\widetilde{\gamma _{+}}$. These $%
L_{0}$ and $\widetilde{\gamma _{+}}$ are called an $L_{0}$-splitting of the $%
L_{+}$.\newline
Because all the knots we consider in this section are non-oriented, the
added segments $\overline{A^{\prime }C^{\prime }}$, $\overline{D^{\prime
}B^{\prime }}$, $\overline{AC}$ and $\overline{DB}$ should have no
orientations either. Hence no path-cancellation may take place between these
segments, different from what happened in the last section. The
contributions of these segments to $t^{I\left( L_{+}\right) }$ should be
discussed individually.\newline
Firstly, the $\overline{A^{\prime }C^{\prime }}$ and $\overline{D^{\prime
}B^{\prime }}$ are trivial, because the $\overline{MA^{\prime }C^{\prime }P}%
\oplus \overline{QD^{\prime }B^{\prime }N}$ in Figure \ref{figure_05}(a)
does not contain the double point of $L_{+}$ and is a planar figure. So the
segments $\overline{A^{\prime }C^{\prime }}$ and $\overline{D^{\prime
}B^{\prime }}$ have no contribution to $\int_{\overline{MA^{\prime
}C^{\prime }P}\oplus \overline{QD^{\prime }B^{\prime }N}}A_{i}dx^{i}$, and
hence $\int_{\overline{MA^{\prime }}\oplus \overline{C^{\prime }P}\oplus
\overline{QD^{\prime }}\oplus \overline{B^{\prime }N}}A_{i}dx^{i}$ $=\int_{%
\overline{MA^{\prime }C^{\prime }P}\oplus \overline{QD^{\prime }B^{\prime }N}%
}A_{i}dx^{i}=I\left( L_{0}\right) $.\newline
Secondly, in contrast, in Figure \ref{figure_05}(a) the contributions of $%
\overline{AC}$ and $\overline{DB}$ are non-trivial. The $\widetilde{\gamma
_{+}}$ contains the non-triviality of $L_{+}$ --- the double point, hence as
a stereoscopic figure it cannot be confined in two dimensions. Then the
contributions of the realistic segments $\overline{CD}$ and $\overline{BA}$,
$\int_{\overline{CD}\oplus \overline{BA}}A_{i}dx^{i}$, can only account for
part of the integral over the whole $\widetilde{\gamma _{+}}:\int_{\overline{%
CD}\oplus \overline{BA}}A_{i}dx^{i}$ $=\lambda \oint\nolimits_{\widetilde{%
\gamma _{+}}}A_{i}dx^{i}$ $=\lambda I\left( \widetilde{\gamma _{+}}\right) ,$
where $\lambda $ is a formal ratio constant, $0<\lambda <1$. The $\lambda $
could be evaluated when compared to a concrete model. Since orientations of $%
\widetilde{\gamma _{+}}$ do not affect $\oint_{\widetilde{\gamma _{+}}%
}A_{i}dx^{i}$, the $t^{I\left( \widetilde{\gamma _{+}}\right) }$ is
evaluated as the same as (\ref{selflinknum0}):%
\begin{equation}
t^{I\left( \widetilde{\gamma _{+}}\right) }=t^{\frac{1}{2}}.
\label{gamma+evalu}
\end{equation}%
\newline
Thirdly, then, letting $t^{I\left( L_{+};L_{0}\mathrm{-splitting}\right) }$
be the contribution of $L_{0}$-splitting to $t^{I\left( \widetilde{\gamma
_{+}}\right) }$, we have%
\begin{equation}
t^{I\left( L_{+};L_{0}\mathrm{-splitting}\right) }=t^{\lambda I\left(
\widetilde{\gamma _{+}}\right) }t^{I\left( L_{0}\right) }=t^{\frac{\lambda }{%
2}}t^{I\left( L_{0}\right) }.  \label{L+L0split}
\end{equation}

\item Similarly, to relate $L_{+}$ to $L_{\infty }$ we consider Figure \ref%
{figure_04}(a) and \ref{figure_05}(b). Firstly, as in the above, the $%
\overline{MN}$ of Figure \ref{figure_04}(a) turns to be $\overline{%
MA^{\prime }}\oplus \overline{AB}\oplus \overline{B^{\prime }N}$ of Figure %
\ref{figure_05}(b), and $\overline{PQ}$ of \ref{figure_04}(a) to be $%
\overline{PC^{\prime }}\oplus \overline{CD}\oplus \overline{D^{\prime }Q}$
of \ref{figure_05}(b). Then introduce four imaginary segments $\overline{%
A^{\prime }D^{\prime }}$, $\overline{AD}$, $\overline{CB}$ and $\overline{%
C^{\prime }B^{\prime }}$ into Figure \ref{figure_05}(b), to form an $%
L_{\infty }$ (i.e. $\overline{MA^{\prime }D^{\prime }Q}\oplus \overline{%
PC^{\prime }B^{\prime }N}$) and a $\widetilde{\gamma _{-}}$ (i.e. $\overline{%
ADCBA}$). This is called an $L_{\infty }$-splitting of the $L_{+}$.\newline
Secondly, as above, the $\overline{A^{\prime }D^{\prime }}$ and $\overline{%
C^{\prime }B^{\prime }}$ are trivial and therefore $\int_{\overline{%
MA^{\prime }}\oplus \overline{D^{\prime }Q}\oplus \overline{PC^{\prime }}%
\oplus \overline{B^{\prime }N}}A_{i}dx^{i}$ $=\int_{\overline{MA^{\prime
}D^{\prime }Q}\oplus \overline{PC^{\prime }B^{\prime }N}}A_{i}dx^{i}$ $%
=I\left( L_{\infty }\right) $. The $\overline{AD}$ and $\overline{CB}$ are
non-trivial and hence $\int_{\overline{CD}\oplus \overline{BA}}A_{i}dx^{i}$ $%
=\lambda \int_{\widetilde{\gamma _{-}}}A_{i}dx^{i}$ $=\lambda I\left(
\widetilde{\gamma _{-}}\right) $ $=-\lambda I\left( \widetilde{\gamma _{+}}%
\right) $, where the ratio constant keeps to be $\lambda $ because $%
\widetilde{\gamma _{+}}$ and $\widetilde{\gamma _{-}}$ are mirror-symmetric.%
\newline
Thirdly, letting $t^{I\left( L_{+};L_{\infty }\mathrm{-splitting}\right) }$
be the contribution of $L_{\infty }$-splitting to $t^{I\left( L_{+}\right) }$%
, we have%
\begin{equation}
t^{I\left( L_{+};L_{\infty }\mathrm{-splitting}\right) }=t^{\lambda I\left(
\widetilde{\gamma _{-}}\right) }t^{I\left( L_{\infty }\right) }=t^{-\lambda
I\left( \widetilde{\gamma _{+}}\right) }t^{I\left( L_{\infty }\right) }.
\label{L+Linfsplit}
\end{equation}

\item We deem that the $L_{+}$ represents an interaction between the two
lines $\overline{MN}$ and $\overline{PQ}$ of Figure \ref{figure_04}(a), and
the $L_{0}$- and $L_{\infty }$-splitting are two channels to run this
interaction. Therefore $t^{I\left( L_{+}\right) }$ is expressed as%
\begin{equation}
t^{I\left( L_{+}\right) }=t^{I\left( L_{+};L_{0}\mathrm{-splitting}\right)
}+t^{I\left( L_{+};L_{\infty }\mathrm{-splitting}\right) }=t^{\lambda
I\left( \widetilde{\gamma _{+}}\right) }t^{I\left( L_{0}\right)
}+t^{-\lambda I\left( \widetilde{\gamma _{+}}\right) }t^{I\left( L_{\infty
}\right) }.  \label{L+1}
\end{equation}%
Introducing a constant $a$ as%
\begin{equation}
a\equiv t^{\lambda I\left( \widetilde{\gamma _{+}}\right) }=t^{\frac{\lambda
}{2}},\ \ \ \mathrm{i.e.,}\ \ a^{-1}\equiv t^{\lambda I\left( \widetilde{%
\gamma _{-}}\right) }=t^{-\frac{\lambda }{2}},  \label{non-orientA}
\end{equation}%
(\ref{L+1}) is re-written as $\left\langle L_{+}\right\rangle =a\left\langle
L_{0}\right\rangle +a^{-1}\left\langle L_{\infty }\right\rangle $, which is
the desired first formula of (\ref{Kaufb2}).
\end{itemize}

Similarly, to relate $L_{-}$ to $L_{0}$ and $L_{\infty }$, we consider
Figure \ref{figure_04}(b), \ref{figure_05}(c) and \ref{figure_05}(d).
Firstly, the $\overline{MN}$ of \ref{figure_04}(b) turns to be the $%
\overline{MA^{\prime }}\oplus \overline{AB}\oplus \overline{B^{\prime }N}$
of Figure \ref{figure_05}(c) or \ref{figure_05}(d), and $\overline{PQ}$ of %
\ref{figure_04}(b) to be $\overline{PC^{\prime }}\oplus \overline{CD}\oplus
\overline{D^{\prime }Q}$ of \ref{figure_05}(c) or \ref{figure_05}(d). Then
one introduces $\overline{A^{\prime }C^{\prime }}$, $\overline{AC}$, $%
\overline{DB}$ and $\overline{D^{\prime }B^{\prime }}$ in \ref{figure_05}(c)
to realize an $L_{0}$-splitting of $L_{-}$, and introduces $\overline{%
A^{\prime }D^{\prime }}$, $\overline{AD}$, $\overline{CB}$ and $\overline{%
C^{\prime }B^{\prime }}$ in \ref{figure_05}(d) to realize an $L_{\infty }$%
-splitting of $L_{-}$. Secondly, using a similar analysis for $L_{+}$, we
obtain for $L_{-}$ that%
\begin{equation}
t^{I\left( L_{-};L_{0}\mathrm{-splitting}\right) }=t^{\lambda I\left(
\widetilde{\gamma _{-}}\right) }t^{I\left( L_{0}\right) },\ \ \ \ \ \
t^{I\left( L_{-};L_{\infty }\mathrm{-splitting}\right) }=t^{\lambda I\left(
\widetilde{\gamma _{+}}\right) }t^{I\left( L_{\infty }\right) }.
\label{L-L0Linfsplit}
\end{equation}%
Hence%
\begin{equation}
t^{I\left( L_{-}\right) }=t^{I\left( L_{-};L_{0}\mathrm{-splitting}\right)
}+t^{I\left( L_{-};L_{\infty }\mathrm{-splitting}\right) }=t^{-\lambda
I\left( \widetilde{\gamma _{+}}\right) }t^{I\left( L_{0}\right) }+t^{\lambda
I\left( \widetilde{\gamma _{+}}\right) }t^{I\left( L_{\infty }\right) }.
\end{equation}%
In terms of the constant $a$ we arrive at the second formula of (\ref{Kaufb2}%
): $\left\langle L_{-}\right\rangle =a^{-1}\left\langle L_{0}\right\rangle
+a\left\langle L_{\infty }\right\rangle $. This completes our verification
of the second skein relation (\ref{Kaufb2}) of the Kauffman bracket
polynomial.

\subsection{Skein relation (\protect\ref{Kaufb3})}

\label{subsect3.2}

The third skein relation (\ref{Kaufb3}) is concerned with a union of two
separate realistic components within a link.

\begin{figure}[th]
\centering \includegraphics[scale=0.8]{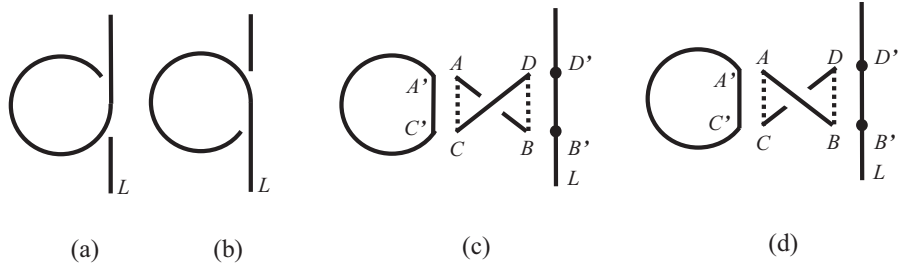}
\caption{Imaginarily adding $\widetilde{\protect\gamma _{+}^{ \prime \prime }%
}$ or $\widetilde{\protect\gamma _{-}^{\prime \prime }}$ to the $L_{c}\sqcup
L$: (a) $\hat{L}_{+}$; (b) $\hat{L}_{-}$; (c) $L_{c}\sqcup L$ with an added $%
\widetilde{\protect\gamma _{+}^{ \prime \prime }}$; (d) $L_{c}\sqcup L$ with
an added $\widetilde{\protect\gamma _{-}^{\prime \prime }}$.}
\label{figure_06}
\end{figure}

Our starting point is to check the following fact for $\left\langle
L_{c}\sqcup L\right\rangle :$%
\begin{equation}
\left\langle L_{c}\sqcup L\right\rangle =a^{-1}\hat{L}_{+}+a\hat{L}_{-},
\label{brack3lemma}
\end{equation}%
where $L_{c}$ is a trivial circle and $L$ an arbitrary link. $\hat{L}_{+}$
comes from adding a degree $\epsilon =+1$ writhe to $L$, as shown in Figure %
\ref{figure_06}(a), and $\hat{L}_{-}$ from adding an $\epsilon =-1$ writhe
to $L$, as shown in Figure \ref{figure_06}(b).

The thought of the last subsection is instructive here for obtaining (\ref%
{brack3lemma}):

\begin{itemize}
\item To relate $\left\langle L_{c}\sqcup L\right\rangle $ to $\hat{L}_{+}$
we consider Figure \ref{figure_06}(c), where an imaginary $\widetilde{\gamma
_{+}^{\phantom{.}\prime \prime }}$ of Figure \ref{figure_03}(c) without
orientation is inserted into $\left\langle L_{c}\sqcup L\right\rangle $. The
$\widetilde{\gamma _{+}^{\phantom{.}\prime \prime }}$ contains two realistic
segments, $\overline{AB}$ and $\overline{CD}$, and two imaginary segments, $%
\overline{AC}$ and $\overline{DB}$. With respect to (\ref{L+L0split}), such
a $\widetilde{\gamma _{+}^{\phantom{.}\prime \prime }}$ has $\int_{%
\widetilde{\gamma _{+}^{\phantom{.}\prime \prime }}}A_{i}dx^{i}=a$. Then,
choosing a trivial segment denoted by $\overline{A^{\prime }C^{\prime }}$ in
the circle $L_{c}$, and a trivial $\overline{D^{\prime }B^{\prime }}$ in the
link $L$, we obtain the union%
\begin{equation}
\hat{L}_{+}=\left( L_{c}\sqcup L\right) \oplus \widetilde{\gamma _{+}^{%
\phantom{.}\prime \prime }},  \label{brack3-L+0}
\end{equation}%
which leads to%
\begin{equation}
L_{c}\sqcup L=\hat{L}_{+}\ominus \widetilde{\gamma _{+}^{\phantom{.}\prime
\prime }},  \label{brack3-L+}
\end{equation}%
where $\ominus $ is the inverse operation of $\oplus $. This is called an $%
\hat{L}_{+}$-insertion of $L_{c}\sqcup L$. Thus, letting $t^{I\left(
L_{c}\sqcup L;\hat{L}_{+}\mathrm{-insertion}\right) }$ be the contribution
of $\hat{L}_{+}$-insertion to $t^{I\left( L_{c}\sqcup L\right) }$, one has%
\begin{equation}
t^{I\left( L_{c}\sqcup L;\hat{L}_{+}\mathrm{-insertion}\right) }=t^{I\left(
\hat{L}_{+}\ominus \widetilde{\gamma _{+}^{\phantom{.}\prime \prime }}%
\right) }=t^{-I\left( \widetilde{\gamma _{+}^{\phantom{.}\prime \prime }}%
\right) }t^{I\left( \hat{L}_{+}\right) }=a^{-1}\left\langle \hat{L}%
_{+}\right\rangle ,  \label{brack3-2}
\end{equation}%
where the sign \textquotedblleft $-$\textquotedblright\ in (\ref{brack3-2})
arises from the operation \textquotedblleft $\ominus $\textquotedblright .

\item Similarly, to relate $\left\langle L_{c}\sqcup L\right\rangle $ to $%
\hat{L}_{-}$ we consider Figure \ref{figure_06}(d), where a $\widetilde{%
\gamma _{-}^{\phantom{.}\prime \prime }}$ of Figure \ref{figure_03}(d) is
inserted into $\left\langle L_{c}\sqcup L\right\rangle $. The $\widetilde{%
\gamma _{-}^{\phantom{.}\prime \prime }}$ containing realistic $\overline{AB}
$ and $\overline{CD}$ and imaginary $\overline{AC}$ and $\overline{DB}$ has $%
\int_{\widetilde{\gamma _{-}^{\phantom{.}\prime \prime }}}A_{i}dx^{i}=-a$.
Then one has the union%
\begin{equation}
\hat{L}_{-}=\left( L_{c}\sqcup L\right) \oplus \widetilde{\gamma _{-}^{%
\phantom{.}\prime \prime }},  \label{brack3-L-0}
\end{equation}%
and thus%
\begin{equation}
L_{c}\sqcup L=\hat{L}_{-}\ominus \widetilde{\gamma _{-}^{\phantom{.}\prime
\prime }}.  \label{brack3-L-}
\end{equation}%
This is called an $\hat{L}_{-}$-insertion of $L_{c}\sqcup L$. Then, letting $%
t^{I\left( L_{c}\sqcup L;\hat{L}_{-}\mathrm{-insertion}\right) }$ be the
contribution of $\hat{L}_{-}$-insertion to $t^{I\left( L_{c}\sqcup L\right)
} $, one has%
\begin{equation}
t^{I\left( L_{c}\sqcup L;\hat{L}_{-}\mathrm{-insertion}\right) }=t^{I\left(
\hat{L}_{-}\ominus \widetilde{\gamma _{-}^{\phantom{.}\prime \prime }}%
\right) }=t^{-I\left( \widetilde{\gamma _{-}^{\phantom{.}\prime \prime }}%
\right) }t^{I\left( \hat{L}_{-}\right) }=a\left\langle \hat{L}%
_{-}\right\rangle .
\end{equation}

\item We deem that in Figure \ref{figure_06}(a) at the double point occurs
the self-interaction of the vortex line $\hat{L}_{+}$, which has $\left(
L_{c}\sqcup L\right) $ as one of its interaction channels. Similarly, in
Figure \ref{figure_06}(b) there occurs the self-interaction of $\hat{L}_{-}$
which has $\left( L_{c}\sqcup L\right) $ as an interaction channel. These
imply $\left\langle L_{c}\sqcup L\right\rangle $ receives contributions from
both Figure \ref{figure_06}(a) and \ref{figure_06}(b):%
\begin{equation}
\left\langle L_{c}\sqcup L\right\rangle =t^{I\left( L_{c}\sqcup L;\hat{L}_{+}%
\mathrm{-insertion}\right) }+t^{I\left( L_{c}\sqcup L;\hat{L}_{-}\mathrm{%
-insertion}\right) }=a^{-1}\left\langle \hat{L}_{+}\right\rangle
+a\left\langle \hat{L}_{-}\right\rangle .  \label{brack3lemma-2}
\end{equation}%
(\ref{brack3lemma-2}) gives the required expression (\ref{brack3lemma}) for $%
\left\langle L_{c}\sqcup L\right\rangle $.
\end{itemize}

Then, on the other hand, according to the skein relation (\ref{Kaufb2}), $%
\left\langle \hat{L}_{+}\right\rangle $ and $\left\langle \hat{L}%
_{-}\right\rangle $ can also be obtained from $\left\langle
L_{+}\right\rangle $ and $\left\langle L_{-}\right\rangle $ as%
\begin{equation}
\left\langle \hat{L}_{+}\right\rangle =a\left\langle L_{c}\sqcup
L\right\rangle +a^{-1}\left\langle L\right\rangle ,\ \ \ \ \ \ \ \ \ \ \
\left\langle \hat{L}_{-}\right\rangle =a^{-1}\left\langle L_{c}\sqcup
L\right\rangle +a\left\langle L\right\rangle .  \label{brack3-skein2}
\end{equation}%
Thus substituting (\ref{brack3-skein2}) into (\ref{brack3lemma-2}) we
precisely acquire%
\begin{equation}
\left\langle L_{c}\sqcup L\right\rangle =-\left( a^{2}+a^{-2}\right)
\left\langle L\right\rangle .  \label{Kaufb3-2}
\end{equation}%
(\ref{Kaufb3-2}) gives the third skein relation (\ref{Kaufb3}) of the
Kauffman bracket polynomial. We address that the sign \textquotedblleft $-$%
\textquotedblright\ in the RHS of (\ref{Kaufb3-2}) should be understood as a
consequence of the above algebraic deduction of (\ref{Kaufb3-2}).

A point should be stressed. The Kauffman bracket polynomial of a single loop
$\gamma _{+}$ is $\left( -a^{3}\right) $, obtained from splitting the double
point of $\gamma _{+}$ and using the skein relations (\ref{Kaufb2}) and (\ref%
{Kaufb3}). It is incorrect to directly use (\ref{gamma+evalu}) to evaluate $%
\left\langle \gamma _{+}\right\rangle =t^{I\left( \widetilde{\gamma _{+}}%
\right) }=a$, because in the context of (\ref{gamma+evalu}) the $\widetilde{%
\gamma _{+}}$ is an imaginary writhe rather than a realistic component.
Similarly, for a single $\gamma _{-}$ its bracket polynomial reads $\left(
-a^{-3}\right) $.

\subsection{Example: trefoil knot}

\label{subsect3.3}

\begin{figure}[th]
\centering \includegraphics[scale=0.8]{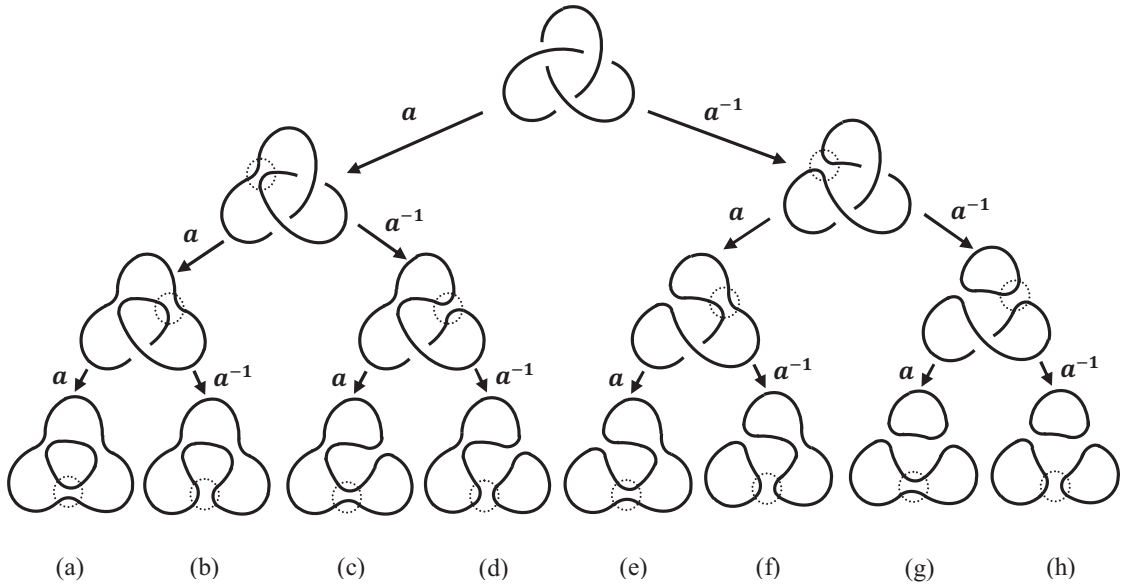}
\caption{Computation of the Kauffman bracket polynomial of a right-handed
trefoil knot. Top row: the right-handed trefoil knot. Bottom row: (a) -- (h)
are the eight statuses obtained after splitting the three double points of
the trefoil knot.}
\label{figure_07}
\end{figure}

As an example, let us compute the Kauffman bracket polynomial of a
right-handed trefoil knot in Figure \ref{figure_07} in the light of the
skein relations (\ref{Kaufb1}) to (\ref{Kaufb3}).

Observe the three double points of the trefoil knot. Without loss of
generality we regard each double point as an \textquotedblleft $L_{+}$%
\textquotedblright -crossing, then the point has two kinds of splitting, the
$L_{0}$- and $L_{\infty }$-splitting, which respectively contribute an
\textquotedblleft $a$\textquotedblright\ and an \textquotedblleft $a^{-1}$%
\textquotedblright\ to the $\left\langle L_{+}\right\rangle $, according to (%
\ref{Kaufb2}). Thus, splitting the three double points one by one, as shown
in Figure \ref{figure_07}, we arrive at the eight completely-split figures
shown in Figure \ref{figure_07}(a) -- \ref{figure_07}(h). Each figure is
called a status. Their respective polynomials are computed as follows:

\begin{itemize}
\item Status \ref{figure_07}(a) comes from the original trefoil knot through
three $L_{0}$-splittings which contribute $a^{3}$, according to (\ref{Kaufb2}%
); Status \ref{figure_07}(a) contains two separate trivial circles which
contribute one $\left( -a^{2}-a^{-2}\right) $, according to (\ref{Kaufb3})
and (\ref{Kaufb1}). Therefore the bracket polynomial of Status \ref%
{figure_07}(a) reads: $-a^{3}\left( a^{2}+a^{-2}\right) $.

\item Status \ref{figure_07}(b) comes through two $L_{0}$-splittings and one
$L_{\infty }$-splitting which totally contribute $a^{2}a^{-1}=a$, according
to (\ref{Kaufb2}); Status \ref{figure_07}(b) contains one circle which
contributes a $1$, according to (\ref{Kaufb1}). Therefore the polynomial of
Status \ref{figure_07}(b) reads: $a$.

\item Similarly, the polynomials of Status \ref{figure_07}(c) to \ref%
{figure_07}(h) read:%
\begin{equation*}
\begin{tabular}{lllll}
\ref{figure_07}(c): & $a;$ &  & \ref{figure_07}(d) & $-a^{-1}\left(
a^{2}+a^{-2}\right) ;$ \\
\ref{figure_07}(e): & $a;$ &  & \ref{figure_07}(f): & $-a^{-1}\left(
a^{2}+a^{-2}\right) ;$ \\
\ref{figure_07}(g): & $-a^{-1}\left( a^{2}+a^{-2}\right) ;$ &  & \ref%
{figure_07}(h): & $a^{-3}\left( a^{2}+a^{-2}\right) ^{2}.$%
\end{tabular}%
\end{equation*}
\end{itemize}

\noindent Hence the Kauffman bracket polynomial of the trefoil knot of
Figure \ref{figure_07} is the sum of the polynomials of Status \ref%
{figure_07}(a) -- \ref{figure_07}(h):%
\begin{equation}
\left\langle \mathrm{Right\ handed\ trefoil\ knot}\right\rangle
=-a^{5}-a^{-3}+a^{-7}.
\end{equation}%
Similarly, for a left-handed trefoil knot --- the mirror image of the
right-handed trefoil knot, obtained by changing the crossing situation of
each double point to its inverse crossing --- its bracket polynomial reads: $%
-a^{-5}-a^{3}+a^{7}$.

For a generic link $\mathcal{L}$ its Kauffman bracket polynomial can be
similarly obtained by using the above status model. The result is%
\begin{equation}
\left\langle \mathcal{L}\right\rangle =\sum_{s}a^{\theta _{0}\left( s\right)
}a^{-\theta _{\infty }\left( s\right) }\left[ \left( -a^{2}-a^{-2}\right)
^{\left\vert s\right\vert -1}\right] ,  \label{Kaufstatusmodel}
\end{equation}%
where all the double points of $\mathcal{L}$ have been split and $s$ denotes
one of the statuses. $\theta _{0}\left( s\right) $ refers to the number of $%
L_{0}$-splittings during the splitting procedure of $\mathcal{L}$ towards
obtaining Status $s$, and $\theta _{\infty }\left( s\right) $ refers to the
number of $L_{\infty }$-splittings during the procedure towards obtaining $s$%
. The $\left\vert s\right\vert $ denotes the number of components (namely
separate trivial circles) appearing in Status $s$.

\subsection{Jones polynomial}

\label{subsect3.4}

The Jones polynomial for oriented links can be constructed from the Kauffman
bracket polynomial \cite{Kauffmanbook2}.

\begin{figure}[th]
\centering \includegraphics[scale=0.8]{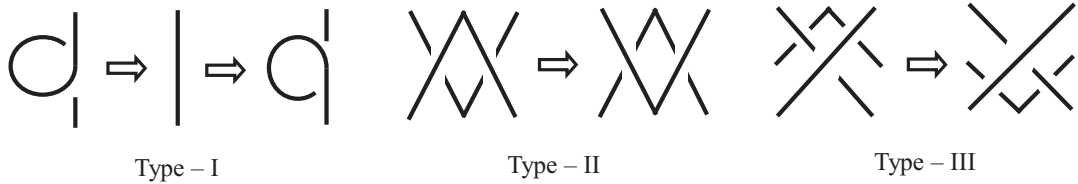}
\caption{Three types of Reidemeister moves. }
\label{figure_08}
\end{figure}

The Jones polynomial is ambient isotopic, namely, it is invariant under all
the three types of Reidemeister moves shown in Figure \ref{figure_08}.
However, the Kauffman bracket polynomial is regularly isotopic, i.e., it is
invariant only under type-II and -III Reidemeister moves, because the
difference between $\left\langle \hat{L}_{\pm }\right\rangle $ and $%
\left\langle \hat{L}\right\rangle $ can be found with respect to (\ref%
{Kaufb2}) and (\ref{Kaufb3}):%
\begin{equation}
\left\langle \hat{L}_{+}\right\rangle =-a^{3}\left\langle \hat{L}%
\right\rangle ,\ \ \ \ \ \ \left\langle \hat{L}_{-}\right\rangle
=-a^{-3}\left\langle \hat{L}\right\rangle ,  \label{brack-typeImoves}
\end{equation}%
which says the Kauffman bracket polynomial is not invariant under type-I
Reidemeister moves. Therefore, in order to construct the Jones polynomial
from the Kauffman bracket polynomial, one should not only endow knots with
orientations, but also modify the bracket polynomial to be invariant under
type-I moves.

From the algebraically topological point of view, the difference between $%
\left\langle \hat{L}_{\pm }\right\rangle $ and $\left\langle \hat{L}%
\right\rangle $ is given by%
\begin{equation}
\left\langle \hat{L}_{+}\right\rangle =\alpha \left\langle \hat{L}%
\right\rangle ,\ \ \ \ \ \ \left\langle \hat{L}_{-}\right\rangle =\alpha
^{-1}\left\langle \hat{L}\right\rangle ,  \label{brack-typeImoves2}
\end{equation}%
where $\alpha $ is a constant caused by adding a degree $\epsilon =1$ writhe
to a link, and $\alpha ^{-1}$ corresponds to the addition of an $\epsilon
=-1 $ writhe. Comparing (\ref{brack-typeImoves}) and (\ref{brack-typeImoves2}%
) one obtains that $\alpha =-a^{3}$. Then, a new polynomial $V\left(
\widehat{\mathcal{L}}\right) $ of $\widehat{\mathcal{L}}$ can be constructed
from the Kauffman bracket polynomial $\left\langle \mathcal{L}\right\rangle $
by compensating the impact of $\alpha $:%
\begin{equation}
V\left( \widehat{\mathcal{L}}\right) \equiv \alpha ^{-w\left( \mathcal{L}%
\right) }\left\langle \mathcal{L}\right\rangle ,  \label{JonesDef}
\end{equation}%
where $\widehat{\mathcal{L}}$ is an oriented link obtained by endowing a
non-oriented link $\mathcal{L}$ with orientations. The $w\left( \mathcal{L}%
\right) $, called the algebraic writhing number of the link $\mathcal{L}$,
is defined as $w\left( \mathcal{L}\right) \equiv \sum_{p}\epsilon \left(
p\right) $, where $p$ denotes all the double points of $\mathcal{L}$, and $%
\epsilon \left( p\right) $ the degree of the point $p$. Now it can be
checked that $V\left( \widehat{\mathcal{L}}\right) $ is an ambient isotopic
polynomial%
\begin{equation}
V\left( \widehat{\hat{L}_{\pm }}\right) =V\left( \widehat{\hat{L}}\right) ,
\end{equation}%
where $w\left( \hat{L}_{\pm }\right) =w\left( \hat{L}\right) \pm 1$ applies.

Eliminating $\left\langle L_{\infty }\right\rangle $ from the two formulae
of (\ref{Kaufb2}), one has $a\left\langle L_{+}\right\rangle
-a^{-1}\left\langle L_{-}\right\rangle =\left( a^{2}-a^{-2}\right)
\left\langle L_{0}\right\rangle $. Replacing $\left\langle
L_{+,-,0}\right\rangle $ with $V\left( \widehat{L_{+,-,0}}\right) $ and
noticing $w\left( L_{\pm }\right) =w\left( L_{0}\right) \pm 1$, one obtains%
\begin{equation}
a^{4}V\left( \widehat{L_{+}}\right) -a^{-4}V\left( \widehat{L_{-}}\right)
=\left( a^{-2}-a^{2}\right) V\left( \widehat{L_{0}}\right) .
\label{JonesPoly1}
\end{equation}%
Then, introducing a constant $\tau =a^{-4}$ for (\ref{JonesPoly1}), and
explicitly writing out (\ref{Kaufb1}), we acquire%
\begin{eqnarray}
&&V\left( L_{c}\right) =1  \label{JonesPoly0} \\
&&\tau ^{-1}V\left( \widehat{L_{+}}\right) -\tau V\left( \widehat{L_{-}}%
\right) =\left( \tau ^{\frac{1}{2}}-\tau ^{-\frac{1}{2}}\right) V\left(
\widehat{L_{0}}\right) .  \label{JonesPoly2}
\end{eqnarray}%
Eqs.(\ref{JonesPoly0}) and (\ref{JonesPoly2}) are recognized to be the
well-known skein relations of the Jones polynomial. Hence $V\left( \widehat{%
\mathcal{L}}\right) $ is the desired Jones polynomial for oriented links.

\section{Conclusion and discussion}

\label{Section_4}

In this paper we attempted to establish a direct relationship between the
abelian CS action and link polynomial invariants of knot theory. We
constructed a topological invariant $t^{I\left( \mathcal{L}\right) }$ for a
link $\mathcal{L}$. In Section \ref{Section_2} it was shown that for
oriented knotted vortex lines, $t^{I}$ satisfies the skein relations of the
Kauffman R-polynomial. In Section \ref{Section_3} it was shown that for
un-oriented knotted lines, $t^{I}$ satisfies the skein relations of the
Kauffman bracket polynomial. As an example the bracket polynomials of the
right- and left-handed trefoil knots were computed, and the Jones polynomial
was constructed from the bracket polynomial. Our emphasis was placed on
Section \ref{Section_3}.

A point may be discussed. In Section \ref{Section_1} it was pointed out that
the CS action $I$ can be expressed as $I=\sum_{k}\oint\nolimits_{\xi
_{k}}A_{i}dx^{i}$ and the gauge potential $A_{i}$ has a decomposition $%
A_{i}=\epsilon ^{ab}n^{a}\partial _{i}n^{b}$. Noticing $n^{a}$ is
ill-defined on vortex lines, the Chern-Simons action $I$ contains
indeterminateness. Therefore the use of Eq.(\ref{naredefinitn}) indeed means
choosing a gauge for $I$. One can expect that other different choices of
gauge conditions may yield different integration result, and thus yield
different polynomial invariants for knots.

\section{Acknowledgment}

The author is indebted to Prof. Ruibin Zhang for useful discussions on the
Kauffman polynomials and constant help in research. This work was
financially supported by the USYD Postdoctoral Fellowship of the University
of Sydney, Australia.

\end{document}